\documentclass[conference]{IEEEtran}
\IEEEoverridecommandlockouts
\usepackage{cite}
\usepackage{amsmath,amssymb,amsfonts}
\usepackage{algorithmic}
\usepackage{graphicx}
\usepackage{textcomp}
\usepackage{xcolor}
\usepackage{booktabs}
\usepackage[hidelinks]{hyperref}
\usepackage{url}
\def\BibTeX{{\rm B\kern-.05em{\sc i\kern-.025em b}\kern-.08em
    T\kern-.1667em\lower.7ex\hbox{E}\kern-.125emX}}
\begin{document}

\title{Who Is Talking to the Agent? LLMs in Multi-User 3D Virtual Environments}%

\author{\IEEEauthorblockN{Mohammad Al-Ratrout, Shayla Sharmin, and Roghayeh Leila Barmaki}
\IEEEauthorblockA{University of Delaware}}

\maketitle

\begin{figure*}[t]
  \centering
  \includegraphics[width=\textwidth]{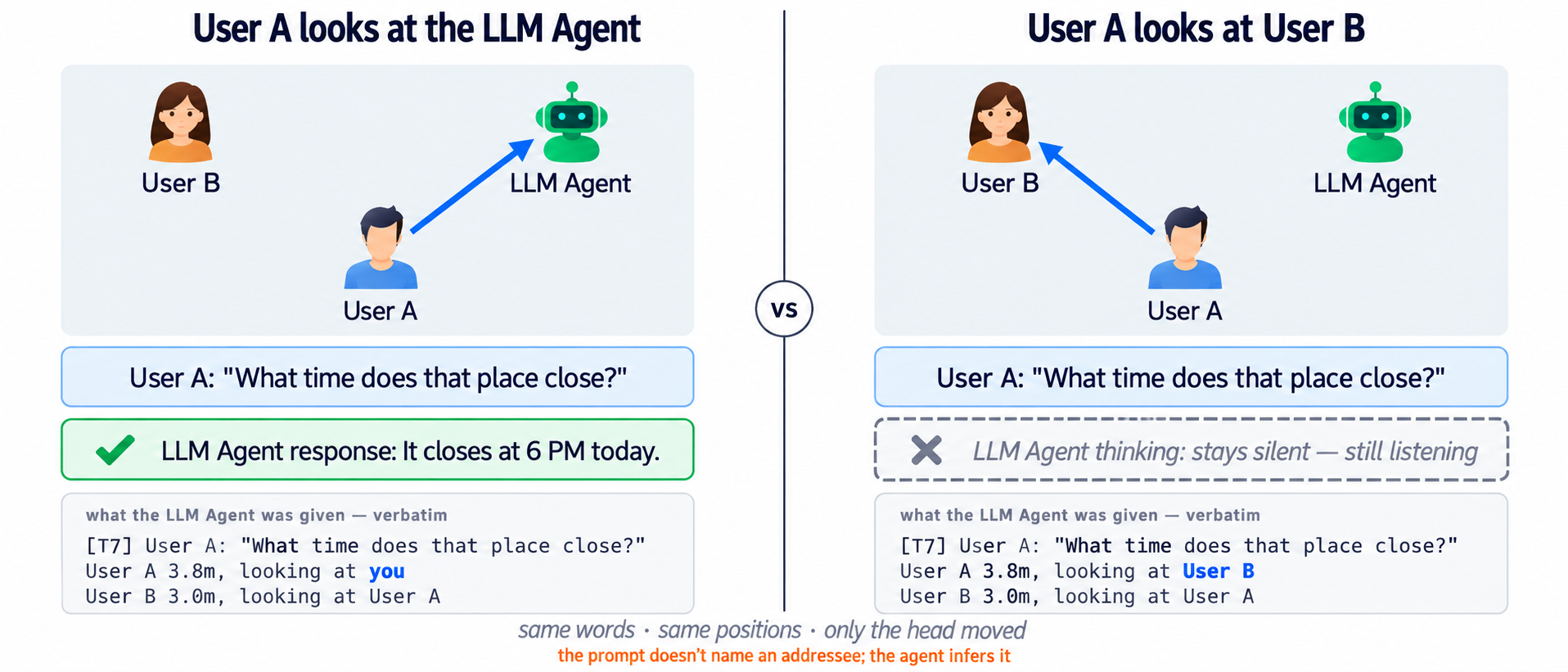}
\caption{The same sentence, twice. User~A asks ``What time does that place close?''
in a shared room with User~B and an LLM agent. Looking at the LLM agent (left) it
answers; looking at User~B (right) it stays silent, still listening. Only the head
direction changed. Beneath each panel is the prompt the agent received, verbatim:
the two differ in a single line, and neither names an addressee. The number after each name is that person's distance from the LLM agent. Across 18,000
decisions, models do not weigh this cue against language; the two models that pass
the baseline addressee check follow it on 85.0--94.2\% of turns where the two
disagree.}
  \label{fig:teaser}
\end{figure*}

\begin{abstract}
When several people share a 3D virtual room with an LLM agent, the agent must
decide not only what to say, but whether an utterance was addressed to it and,
if accessible, what profile information about the others present it may use.
To study both problems, we construct \textsc{LookAway}, a controlled corpus of
40 sessions involving 80 distinct personas and an LLM agent (1,200 turns),
including ambiguous-addressee turns in which speaker orientation agrees or
conflicts with the intended addressee. Across three open-weight large language
models and five conditions varying which profiles the agent sees and whether it
is told where each person faces (18,000 decisions), adding speaker orientation
increased addressee accuracy from 56\% to 99.5\% when orientation was
congruent, but when the speaker faced someone other than the addressee, two of
the models went by where the speaker faced on more than 85\% of those turns.
Warning one model that orientation could be misleading reduced this only
modestly. A browser-based 3D demonstrator shows the effect live: the same
sentence gets an answer when the speaker faces the agent and silence when they
face the other person. Providing both personas' profiles improved responses
about the person being asked about, but also increased the use of profile
attributes not revealed in the shared conversation, reaching 45.3\% of answers
for one model. More context thus improves multi-user interaction but also leads
to oversharing, so shared LLM agents need mechanisms for weighing spatial cues
and controlling when user-specific information enters a response. \href{https://github.com/m7md2216/LLMs-in-Multi-User-3D-Virtual-Environments}{link to our code}.
\end{abstract}


\section{Introduction}
\label{sec:introduction}

Large language model (LLM)-powered agents are increasingly appearing as
embodied conversational partners in shared virtual environments, including
XR systems and commercial metaverse platforms
\cite{cuify2024,hiroi2026}. As these agents move beyond one-to-one
interaction, they must determine not only what to say, but whether an
utterance was addressed to them and whose information to use in responding.
Unlike conventional voice assistants that rely on wake words or explicit
invocation, shared agents must make these decisions within ongoing
multi-party conversation.

Speaker orientation provides a useful spatial cue to addressee intent  (Fig.~\ref{fig:teaser}). In
human multi-party conversation, gaze and head direction are associated with
whom a speaker addresses, and virtual environments can provide avatar
orientation directly rather than estimating it from video. Yet it remains
unclear how LLM-based agents use such information. In a recent benchmark of
recorded three-person conversations, GPT-4o achieved 80.9\% addressee
accuracy against an 80.6\% majority-class baseline, while adding video-estimated gaze
reduced accuracy to 75.2\%~\cite{addressee_bench}. This result cannot
distinguish whether the model used gaze poorly or the estimated signal itself
was unreliable. Such data are also hard to collect in the physical world, where conversations must be recorded and head orientation estimated from video. Virtual environments avoid both problems: every avatar's position and head direction are known exactly, so the same conversation can be replayed with orientation that agrees or conflicts with the intended addressee.

A further difficulty is that in natural conversations, spatial orientation
usually agrees with the intended addressee. Such cases cannot distinguish
integrating spatial information from simply following it. We therefore
developed \textsc{LookAway}, a controlled multi-user conversational corpus in
which speaker orientation can agree or conflict with the assigned addressee.
The corpus also varies the profile information available about two co-present
users, allowing us to examine both addressee decisions and profile use.
We complement the corpus with a browser-based 3D demonstrator that derives
speaker orientation from users' head direction during interaction. Our research questions are:

\textbf{RQ1:} When speaker orientation conflicts with the assigned addressee,
how does an LLM-based agent determine whether it is being addressed?

\textbf{RQ2:} When an agent has access to multiple users' profiles, how does
that information affect what it says about a co-present person?

Our results reveal two patterns. First, the tested agents follow speaker
orientation on more than 85\% of incongruent turns, and explicitly warning
\texttt{qwen3.6} that orientation may be misleading reduces this behavior
only modestly. Second, providing both users' profiles increases responses
about the intended person but also increases the use of profile attributes
that person has not revealed in the conversation, so the agent can tell one person what the other never chose to say. A privacy instruction reduces this, but its strictest wording also makes the agent refuse questions whose answers were already public.

Our contributions are:
\begin{itemize}\itemsep2pt
    \item \textsc{LookAway}, a controlled multi-user conversational corpus
    that independently varies assigned addressee, speaker orientation, and
    profile availability.

    \item controlled evaluations of addressee decisions and multi-user
    profile use across open-weight LLM agents.

    \item a browser-based 3D demonstrator that instantiates the addressee
    setting using users' runtime head direction.
\end{itemize}
\section{Related Work}
\label{sec:related}

\subsection{Addressee Detection in Multi-Party Interaction}

Identifying who is being addressed has long been studied in multi-party
interaction. Prior work combines gaze or head orientation with speech and
conversational context to identify the addressee
\cite{jovanovic2006addressee,addressee_nonverbal2011}. Gaze has also been
used by conversational agents and robots to manage turn-taking and decide
when to respond
\cite{bohus2010facilitating,robotgaze_roles,explainrobot2024}.
These studies establish spatial behavior as a useful cue to addressee intent.

Recent work asks whether language and multimodal models can make the same
decision. On recorded three-person conversations, GPT-4o achieved 80.9\%
addressee accuracy against an 80.6\% majority-class baseline. Its accuracy fell to 75.2\% when
video-estimated gaze was added~\cite{addressee_bench}. Models also lose accuracy on conversational roles when character identities are anonymized~\cite{convstructure2025}, and addressee recognition depends on conversation structure rather than text alone~\cite{hangover2024}. Video-LLMs likewise track characters poorly, often relying on coarse cues such as gender~\cite{videollm2026}.

These results do not show what happens when a spatial cue conflicts with the
assigned addressee. Video-based studies also make it difficult to separate
model behavior from errors in gaze estimation. \textsc{LookAway} addresses
both issues by providing controlled speaker orientation directly and by
including turns in which orientation agrees or conflicts with the assigned
addressee.

This problem is especially relevant for shared agents that listen while
people also talk to one another. Overhearing agents are designed to enter
such conversations only when useful~\cite{overhearing2025}, while systems
such as GroupGPT separately decide when an agent should speak in group
interaction~\cite{groupgpt2026}. Shared virtual environments add a useful
spatial signal because avatar head orientation can be obtained directly from
the environment. LLM-based agents are already being explored in XR and
commercial virtual environments
\cite{cuify2024,xracceptance2025,hiroi2026}.

\subsection{Shared Agents and Multi-User Profiles}

Multi-user agents must also keep information associated with the correct
person. AFA links memories to individual users to reduce cross-user
confusion~\cite{afa2026}. PAChat links speakers in group conversation to
their profiles to personalize responses~\cite{pachat2025}, while
MAP studies how agents handle conflicting preferences across users
\cite{map2025}.

These systems mainly address who is speaking, which information belongs to
which user, or how different users' preferences should be handled. Our
setting adds a different case: one person asks the agent about another
co-present person. We examine whether access to both profiles helps the agent
respond about the person being discussed and whether it also changes what
profile information appears in the response.

\subsection{Information Sharing Across Users}

Access to a person's information does not necessarily make every use of that
information appropriate. Contextual integrity describes privacy in terms of
whether information flows appropriately within a given context
\cite{nissenbaum2004privacy}. Recent LLM work applies this idea to
interactive systems in which models receive information from multiple sources
and may reveal it in an inappropriate context
\cite{mireshghallah2024can}. Other work studies information leakage
between agents, tools, and users
\cite{agentleak2026,magpie2025,agentdam2025,privacyaction2025,multiuser2026}.

We measure whether information available
only through a co-present person's profile appears when another person asks
about them. This separates two questions: whether the agent keeps users'
identities distinct, and whether it limits its use of information that was
not revealed in the shared conversation.

Prior work uses profile-based automatic evaluation for personalized
responses \cite{pref2025,semconstraint2026,persobench2024}.
Because our measures rely partly on automatic extraction and scoring,
we additionally compare them with human judgments
(\S\ref{sec:validation}).

\section{\textsc{LookAway}: Corpus and 3D Demonstrator}
\label{sec:lookaway}

This section describes \textsc{LookAway}, a controlled multi-user conversational corpus, and a separate browser-based 3D demonstrator. The corpus contains 40 sessions, each with
two personas and an LLM agent, totaling 1,200 turns. Conversations and spatial
states are synthetically constructed; the demonstrator instantiates the same
setting interactively using runtime head orientation.

\subsection{Persona and Profile Construction}

The 40 sessions use 80 distinct personas drawn from the Personalized Agent
chaT (PAT) dataset~\cite{afa2026}. PAT profiles include demographics, career,
personality, motivations and values, decision-making style, preferences, and
emotional triggers. We use profile attributes and scenario names but exclude
scenario-specific scenes because some contain profile information intended to
remain unrevealed.
We exclude 13 personas with fantastical content and personas with fewer than
eight usable attributes. The remaining personas are paired based on
contrasting attributes, with no persona reused across sessions. The 12 PAT
scenarios rotate across the 40 sessions.

Each persona has eight attributes: four \textbf{revealed} in the conversation
and four \textbf{latent}, available only through the profile. Revealed
attributes are selected first. Each latent attribute must have cosine
similarity below 0.45 to every revealed attribute using MiniLM embeddings,
reducing direct overlap between the two sets. Ties are resolved using contrast
coverage, category diversity, and revealed--latent similarity.

The final corpus contains 640 attributes: 435 from PAT and 205 authored for
\textsc{LookAway}. Authored attributes come from a fixed bank of opposing traits covering
12 concrete dimensions: spice, caffeine, transport, noise, planning ahead,
alcohol, exercise intensity, diet, budget, temperature, crowds, and schedule.
For each persona pair, three opposing dimensions are selected using a seeded
procedure, producing 120 contrasts. These opposing attributes make
wrong-person attribution observable.

\subsection{Conversation and Addressee Design}

Each session contains 30 turns: six clearly addressed to the agent, sixteen
clearly addressed to the other person, and eight ambiguous-addressee turns
(``traps''). The six agent-addressed turns include three name-addressed and
three task-request turns. The sixteen person-addressed turns include four
name-addressed, six private-context, and six side-talk turns. Of the eight
traps, four are assigned to the agent and four to the other person.
Turn order is generated by a seeded program. Speakers alternate; one
agent-directed turn occurs within each block of three; the first turn is not
agent-directed, and no more than two agent-directed turns occur consecutively.

Across the corpus, this yields 240 clearly agent-addressed turns, 640 clearly
person-addressed turns, and 320 traps. The 880 clear turns check whether the
agent responds when addressed and remains silent otherwise. The traps test
addressee decisions when wording alone does not identify the listener. Half
are assigned to the agent and half to the other person, so always responding
or always remaining silent gives 50\% accuracy.
Trap utterances contain no persona names, assistant references,
assistant-specific requests, or private history that uniquely identifies the
listener. These constraints are checked automatically, and failed utterances
are regenerated up to three times. Sixteen traps required regeneration.

The assigned addressee is set by the conversation plan before generation.
The utterance generator (\texttt{gemma4:31b}; Section~\ref{sec:generation}) is
told whom the utterance targets but is instructed to keep that target unclear
from the wording. Thus, the evaluation uses the
corpus-assigned addressee rather than an addressee inferred from the text.

\subsection{Spatial-Cue Manipulation}

The corpus uses synthetic spatial states. The agent is placed at the origin,
and the two personas are placed at seeded random positions 2--6\,m away, with
at least 0.6 radians of angular separation. Positions remain fixed within a
session; persona-to-agent distances range from 2.1 to 6.0\,m
($M=4.06$\,m).

Speaker orientation is set separately for each turn. On congruent turns, the
speaker faces the assigned addressee; on incongruent turns, the speaker faces
the other co-present entity. We therefore refer to this signal as
\emph{speaker orientation} or \emph{head direction}, not measured eye gaze.

Of the 320 traps, 200 are congruent and 120 are incongruent. Each session has
three incongruent traps, limiting cue conflict to 10\% of its 30 turns. These
traps are selected using a seeded random procedure. Across sessions, the
direction of conflict is balanced: 60 turns are assigned to the agent while
the speaker faces the other person, and 60 reverse this relationship. 
Congruent turns test whether the agent can use spatial information when it
agrees with the assigned addressee; incongruent turns reveal whether the
agent follows that information when the two disagree.

\subsection{Experimental Conditions}

\begin{figure}[t]
  \centering
  \includegraphics[width=\columnwidth]{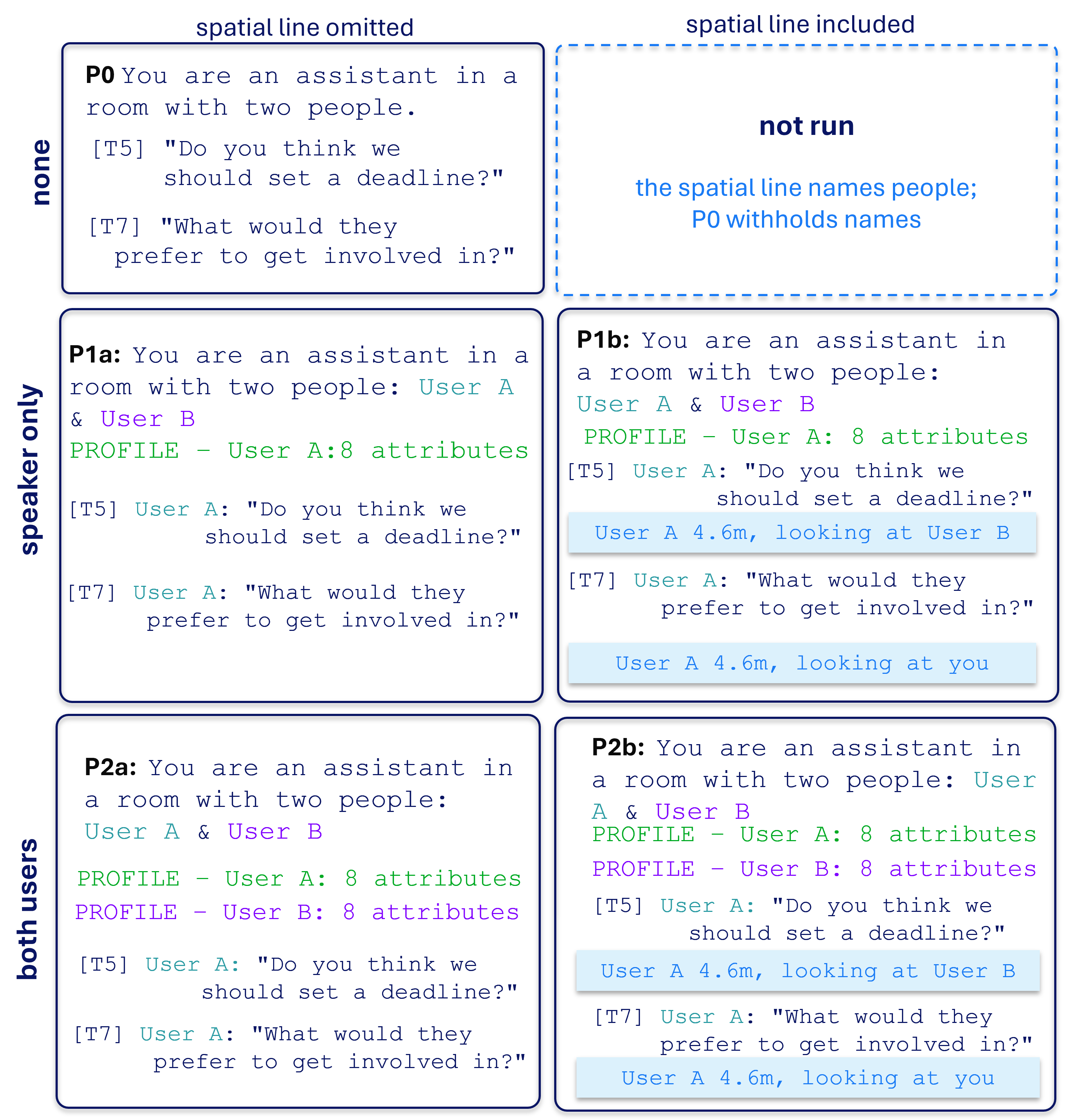}
  \caption{\textbf{What each condition puts in the prompt.} Fragments are taken
  verbatim from the rendering code, abbreviated only in how many turns are
  shown. Tinted lines mark what a cell adds: names and the speaker's profile
  (green), the partner's profile (purple), the spatial line (blue). P1a to P1b, and P2a to P2b, each add exactly one line per turn and nothing else. \textbf{There is no P0 with
  spatial context, because the spatial line names people and P0 withholds
  names.}}
  \label{fig:conditions}
\end{figure}

The same 1,200 turns are evaluated under five conditions
(Fig.~\ref{fig:conditions}):

\begin{itemize}\itemsep1pt
    \item \textbf{P0}: no persona names, profiles, or spatial information;
    \item \textbf{P1a}: names and the current speaker's profile;
    \item \textbf{P1b}: P1a plus spatial information;
    \item \textbf{P2a}: names and both personas' profiles; and
    \item \textbf{P2b}: P2a plus spatial information.
\end{itemize}
P0 provides a no-profile, no-spatial baseline. In Experiment~1, it measures
addressee decisions without identity or spatial information. In Experiment~2,
it provides baselines for identifying the person being discussed and for
apparent disclosure when no profile information is available.
In P1b and P2b, each turn includes a spatial line such as:

\begin{quote}
\texttt{Tom 4.6m, looking at you | Robin 5.5m, looking at Tom}
\end{quote}

Distance is measured from the agent to each persona and remains fixed within
a session; orientation can change each turn. Distance is included as part of
the spatial observation but is not manipulated or analyzed. P0 has no spatial
counterpart because the spatial line identifies personas by name.

The prompt never provides the assigned addressee or indicates whether a turn
is ambiguous, congruent, or incongruent. Each decision is a separate,
stateless request containing only that condition's conversation up to the
current turn; no model memory carries across conditions.

\subsection{Conversation Generation}
\label{sec:generation}
A seeded rule-based program constructs the conversation plan, including turn
type, assigned addressee, and speaker orientation. \texttt{gemma4:31b} then
generates each persona utterance.
The generator receives the speaker's name and four revealed attributes, the
partner's name, scenario, conversation so far, and the instruction for the
current turn. It does not receive latent attributes, the partner's profile,
speaker orientation, or turn-type labels. Generation uses temperature 0.8,
a 90-token output limit, and fixed seeds. Private-context memories are
generated separately with \texttt{granite4.1:8b} using only revealed
attributes.

Trap and cross-reference turns are checked automatically against their
generation constraints. Failed utterances are regenerated with feedback for
up to three attempts. Sixteen of the 1,200 turns required regeneration, all
of them traps; three required all three attempts.

\subsection{Models and Inference Setup}

We evaluate three open-weight agents:
\texttt{qwen3.6:27b}, \texttt{mistral-small3.2:24b}, and
\texttt{nemotron3:33b}. They were selected as models from three vendors that
could run with 4-bit quantization on a single 24\,GB GPU; they are not intended
to represent frontier models. All models were served through Ollama.

The installed models are 27.8B, 24.0B, and 33.0B parameters, respectively,
using Q4\_K\_M quantization.

All agents use temperature 0 and an 8,192-token context window. Addressee
decisions are limited to 40 output tokens and generated replies to 200.
Reasoning mode is disabled. Outputs use JSON with up to four parsing retries;
all 18,000 addressee decisions were successfully parsed.

Five conditions applied to 1,200 turns produce 6,000 decisions per agent and
18,000 decisions overall (Table~\ref{tab:corpus}).

\begin{table}[t]
\centering
\caption{\textsc{LookAway} at a glance. The 120 incongruent traps are the
turns in which speaker orientation conflicts with the assigned addressee.}
\label{tab:corpus}
\begin{tabular}{@{}lr@{}}
\toprule
sessions (two people each) & 40 \\
turns & 1{,}200 \\
\quad clearly to the agent & 240 \\
\quad clearly to the other person & 640 \\
\quad \textbf{traps} & \textbf{320} \\
\qquad congruent / \textbf{incongruent} & 200 / \textbf{120} \\
questions about the other person & 120 \\
attributes per person (revealed / latent) & 4 / 4 \\
conditions & 5 \\
decisions per agent & 6{,}000 \\
\bottomrule
\end{tabular}
\end{table}

\begin{figure*}[t]
  \centering
  \includegraphics[width=1.5\columnwidth]{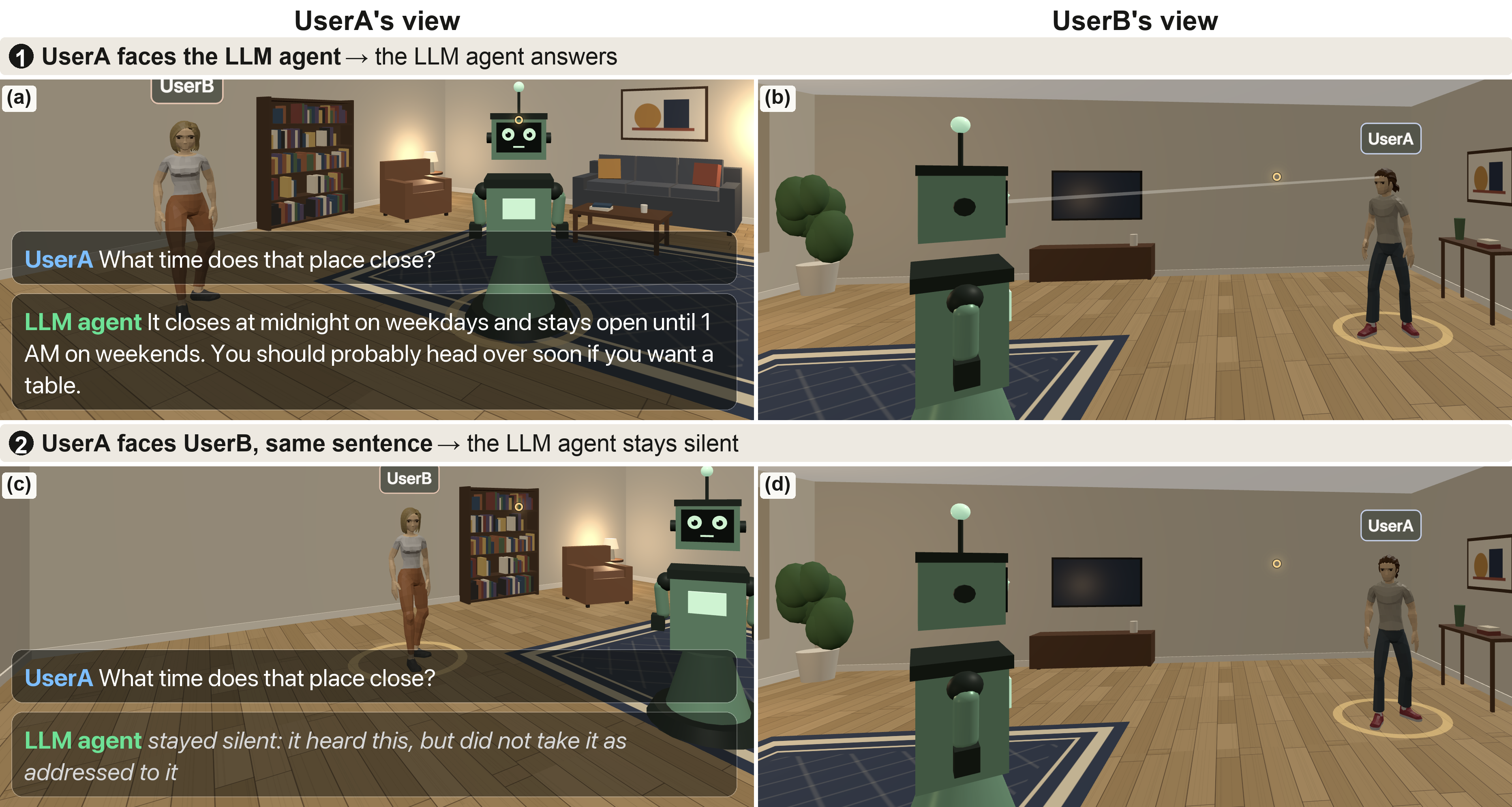}
  \caption{The demonstrator, running Experiment~1's own decision code, seen from
  both people at once. UserA asks the same question twice from the same
  position; only the head direction differs, and the LLM agent is never told who
  is being addressed. The amber ring marks the entity inside the 22\textdegree{} head-direction cone; in UserB's window, UserA's head direction is drawn as a beam, landing on the agent in (b) and absent in (d).}
  \label{fig:demo}
\end{figure*}

\subsection{3D Virtual-Room Demonstrator}
\label{sec:demo}

We also implement the addressee setting in a browser-based 3D room using
Three.js r160 and WebGL. Two users can move, change their viewing direction,
and communicate by typing or push-to-talk. A Python WebSocket server
synchronizes their positions and head directions and runs the agent
server-side.

The demonstrator derives orientation from users' runtime
head direction. An entity is a candidate target when it falls within
22\textdegree{} of the user's facing direction. This threshold is an
implementation choice; it forms a 44\textdegree{}-wide cone and separates the
two targets in the demonstrator layout. If multiple entities fall within the
threshold, the closest in angle is selected; if none does, the observation
records ``not looking at anyone.''

The demonstrator uses \texttt{qwen3.6} with the P1b prompt from Experiment~1, with each profile left empty. The assigned addressee is never supplied to
the model. Figure~\ref{fig:demo} shows the same ambiguous utterance from the
same position: the agent responds when the user faces it and remains silent
when the user faces the other person.
\section{Experiment 1: Addressee Detection Under Conflicting Cues}
\label{sec:exp1}

Experiment~1 examines how an LLM agent decides whether an utterance is
addressed to it. We evaluate \texttt{qwen3.6:27b},
\texttt{mistral-small3.2:24b}, and \texttt{nemotron3:33b}. For each turn, the
model either responds or remains silent. We first test clear addressee turns
and then examine ambiguous turns where speaker orientation agrees or conflicts
with the assigned addressee.

Conditions with and without spatial information use the same turns, so we
compare them using two-sided exact McNemar tests. All eight main spatial comparisons remain significant after Holm correction ($p<0.001$).

\subsection{Baseline Addressee Detection}

We first test the 880 clear turns (240 addressed to the agent and 640 to the other person).
\texttt{qwen3.6} and \texttt{mistral-small3.2} correctly classify 99.5\% and
98.6\% of these turns. \texttt{nemotron3} responds to every agent-addressed
turn but also responds to 61.9\% of turns clearly addressed to the other
person. For example, it often responds when Tom says directly to Robin,
\emph{``Fair enough, Robin, let's lock down those milestones.''}

Because \texttt{nemotron3} does not reliably remain silent when another person
is addressed, the remaining analyses focus on \texttt{qwen3.6} and
\texttt{mistral-small3.2}.

\subsection{Effect of Spatial Context}

Table~\ref{tab:exp1} compares conditions without spatial information
(P1a, P2a) with their spatial counterparts (P1b, P2b). Within each pair,
the prompts are identical except for the line containing the two personas'
distances and orientations.

\begin{table*}[t]
\centering
\caption{\textbf{Effect of adding spatial information on addressee decisions.}
Each accuracy cell reports performance without spatial information
$\rightarrow$ with spatial information. Congruent turns are those in which
speaker orientation is directed toward the assigned addressee; incongruent
turns are those in which it is directed toward the other co-present entity.
Precision and recall are computed over all 1,200 turns in each condition.
P0 is the no-profile, no-spatial baseline; it has no spatial counterpart, so
its cells report a single value and it has no cue-following rate.}
\label{tab:exp1}
\begin{tabular}{@{}llccccc@{}}
\toprule
 & & congruent traps & incongruent traps & follows spatial cue & precision & recall \\
 & & ($n=200$) & ($n=120$) & (incongruent) & & \\
\midrule
\texttt{qwen3.6} & P0 & 58.0 & 53.3 & --- & 97.8 & 65.2 \\
                 & P1 & 56.0 $\rightarrow$ 99.5 & 55.0 $\rightarrow$ \phantom{0}9.2 & 90.8 & 97.7 $\rightarrow$ 86.5 & 64.5 $\rightarrow$ 84.8 \\
                 & P2 & 55.0 $\rightarrow$ 99.0 & 55.8 $\rightarrow$ \phantom{0}5.8 & 94.2 & 97.3 $\rightarrow$ 85.4 & 64.2 $\rightarrow$ 84.5 \\
\midrule
\texttt{mistral} & P0 & 62.5 & 66.7 & --- & 89.1 & 75.8 \\
                 & P1 & 70.0 $\rightarrow$ 96.5 & 62.5 $\rightarrow$ 11.7 & 88.3 & 87.1 $\rightarrow$ 83.0 & 79.2 $\rightarrow$ 86.5 \\
                 & P2 & 68.5 $\rightarrow$ 96.0 & 60.8 $\rightarrow$ 15.0 & 85.0 & 89.7 $\rightarrow$ 85.1 & 76.5 $\rightarrow$ 85.5 \\
\bottomrule
\end{tabular}
\end{table*}

Half of each trap type is assigned to the agent and half to the other person
(100/100 congruent, 60/60 incongruent), so always answering, never answering,
or guessing all score 50\%.

When orientation agrees with the assigned addressee, spatial information
greatly improves accuracy. In P1, \texttt{qwen3.6} increases from 56.0\% to
99.5\%, and \texttt{mistral} from 70.0\% to 96.5\%
($p<0.001$ for both). P2 shows the same pattern
(Table~\ref{tab:exp1}).

When orientation conflicts with the assigned addressee, the pattern reverses.
In P1, accuracy falls from 55.0\% to 9.2\% for \texttt{qwen3.6} and from
62.5\% to 11.7\% for \texttt{mistral} ($p<0.001$). In P2, it falls from
55.8\% to 5.8\% and from 60.8\% to 15.0\%, respectively. All four values are
well below the 50\% baseline.

For \texttt{qwen3.6}, adding spatial information changes 55 incongruent turns
from correct to incorrect and none from incorrect to correct. For example,
Tom asks, \emph{``What are the most critical success factors for this
project?''} on a turn assigned to the agent while facing Robin. Without the
spatial line, the agent often responds; when told that Tom faces Robin, it
remains silent.

\subsection{Cue-Following Behavior}

Because trap wording is intentionally ambiguous, we do not treat the words as
pointing to a different addressee. On each incongruent turn, we instead check
whether the model's decision agrees with the spatial cue or with the assigned
addressee. These point to different targets by design.

\texttt{qwen3.6} follows the spatial cue on 90.8\% of incongruent turns in P1b
and 94.2\% in P2b. The corresponding rates for \texttt{mistral} are 88.3\%
and 85.0\% (Table~\ref{tab:exp1}). Thus, across both profile conditions, the
models follow the spatial cue on 85.0--94.2\% of incongruent turns.

\subsection{Precision--Recall Trade-off}

Precision and recall are computed over all 1,200 turns in each condition,
including 400 assigned to the agent. Precision measures how often the agent's
responses occur on turns assigned to it; recall measures how often it responds
when it is the assigned addressee.

For \texttt{qwen3.6} in P1, adding spatial information increases recall from
64.5\% to 84.8\% but reduces precision from 97.7\% to 86.5\%. P2 shows a
similar pattern. \texttt{mistral} shows the same general shift, but with
smaller changes (Table~\ref{tab:exp1}). Thus, spatial information makes the
models respond to more agent-addressed turns, but also to more turns assigned
to the other person.

\subsection{Robustness to Prompt Warnings}

We next test whether explicitly warning the model about unreliable orientation
reduces its reliance on the spatial cue. This analysis uses
\texttt{qwen3.6}, the model used in the demonstrator, and all 320 P1b traps.
It was not run with \texttt{mistral-small3.2}, so we treat it as a
Qwen-specific check.

The \emph{warned} prompt states that orientation is
\emph{``a cue, not a guarantee \ldots\ often informative but sometimes
wrong.''} The \emph{adversarial} prompt additionally states that people may
look away from the person they are addressing and asks the model to rely on
the conversation when the cues conflict.

\begin{table}[t]
\centering
\caption{\textbf{Effect of warning \texttt{qwen3.6} that orientation may be
misleading (P1b).} Cells are accuracy (\%) against the assigned addressee.
\emph{No warning} is the P1b result from Table~\ref{tab:exp1}. \emph{Follows
spatial cue} is the share of incongruent traps on which the decision matches
the entity the speaker faces (100 minus incongruent accuracy).}
\label{tab:ablation}
\begin{tabular}{@{}lccc@{}}
\toprule
prompt & congruent & incongruent & follows spatial cue \\
 & ($n=200$) & ($n=120$) & (incongruent) \\
\midrule
no warning  & 99.5 & \phantom{0}9.2 & 90.8 \\
warned      & 96.0 & 15.0 & 85.0 \\
adversarial & 91.5 & 18.3 & 81.7 \\
\bottomrule
\end{tabular}
\end{table}

Both warnings reduce, but do not remove, reliance on the spatial cue
(Table~\ref{tab:ablation}). Spatial-cue following falls from 90.8\% to
85.0\% and 81.7\%, while incongruent accuracy rises from 9.2\% to 15.0\%
and 18.3\%. At the same time, congruent accuracy falls from 99.5\% to
96.0\% and 91.5\%.
\section{Experiment 2: Profile Use in Responses About Co-Present Users}
\label{sec:exp2}

Experiment~2 examines what an agent says when one person asks about another.
We analyze the 120 cross-reference turns, such as Emery asking,
\emph{``What does Robin like to do in her free time?''} We test whether
profile access helps the LLM agent respond about the person being discussed and
whether it uses profile information that person has not revealed in the
conversation.

We analyze \texttt{qwen3.6} and \texttt{mistral-small3.2};
\texttt{nemotron3} is excluded because it failed the baseline addressee check
in Experiment~1.

We measure attribute use with an LLM judge~\cite{zheng2023judging, liu2023g}, which we call the extractor: \texttt{deepseek-r1:32b} receives one reply and the list of that pair's attributes, and returns which listed attributes the reply states. This is a closed-list matching task rather than an open-ended rating. The judge differs from both evaluated agents, avoiding direct same-model self-evaluation~\cite{panickssery2024llm}. Its output gives the right-person, wrong-person, disclosure, and mix-up measures defined below; Section~\ref{sec:validation} reports human agreement.

The extractor failed on 177 replies: 102 of 2,000 \texttt{qwen3.6} replies
(5.1\%) and 75 of 2,000 \texttt{mistral} replies (3.8\%). Content measures
use the successfully extracted replies; counting every missing reply in the
least favourable direction does not reverse any comparison in
Table~\ref{tab:exp2}.

Adding the spatial line produced no response-content pattern that repeated
across models and profile conditions. We therefore combine P1a and P1b as P1,
and P2a and P2b as P2 for the main results. The separate spatial comparisons
are reported at the end of this experiment.

\subsection{Response Measures}

Revealed attributes appear in the conversation, whereas latent attributes are
available only through the profile. We define \emph{disclosure} as a response
that states at least one latent attribute of the person being discussed. This
does not mean that every latent attribute is private.

A \emph{right-person response} states at least one attribute of the person
being discussed. A \emph{wrong-person response} states an attribute of the
questioner but none of the other person's attributes. A \emph{mix-up}
attributes one person's fact to the other. Right- and wrong-person responses
cannot overlap. Disclosure can overlap with a right-person response, and in
our data every detected disclosure does.

Response quality is rated from 1 to 5 by \texttt{deepseek-r1:32b}: 5 indicates
clear use of the person's attributes, 3 a reasonable but generic response,
and 1 a generic, evasive, or wrong-person response. Every third reply is
rated; refusals are treated as evasive.

\begin{table}[t]
\centering
\caption{Responses when one persona asks the LLM agent about the other.
``Right person'' indicates use of the target person's attributes; ``wrong
person'' indicates use of the questioner's attributes without the target's;
``disclosure'' indicates use of a latent target-person attribute.
Quality is judged from 1 to 5.}
\label{tab:exp2}
\setlength{\tabcolsep}{3.5pt}
\begin{tabular}{@{}llcccc@{}}
\toprule
 & profiles held & right person & wrong person & disclosure & quality \\
\midrule
\texttt{qwen3.6} & none    & 12.6\% & \phantom{0}0.9\% & \phantom{0}5.4\% & 2.26 \\
                 & speaker & 14.5\% & 29.8\%           & \phantom{0}3.3\% & 2.53 \\
                 & both    & 56.2\% & \phantom{0}6.5\% & 45.3\%           & 2.88 \\
\midrule
\texttt{mistral} & none    & 11.5\% & \phantom{0}2.7\% & \phantom{0}3.9\% & 2.37 \\
                 & speaker & 14.0\% & 28.8\%           & \phantom{0}1.4\% & 2.41 \\
                 & both    & 39.1\% & \phantom{0}8.6\% & 25.9\%           & 2.78 \\
\bottomrule
\end{tabular}
\end{table}

Because P0 provides no profile, its 5.4\% (\texttt{qwen3.6}) and 3.9\%
(\texttt{mistral}) disclosure rates cannot come from profile access. They
represent responses that happen to match latent attributes.

\subsection{Identity Separation}

Across 1,939 successfully extracted replies about the speaker, neither model
attributes the other person's facts to the speaker. These include 965
\texttt{qwen3.6} and 974 \texttt{mistral} replies.

Across all replies, \texttt{qwen3.6} produces no identity mix-ups.
\texttt{mistral} produces five mixed-up attributes in four cross-reference
replies, all when both profiles are available. Thus, identity mix-ups are
rare and do not explain the disclosure results.

\subsection{Effect of Profile Access}

Making both profiles available increases disclosure from 3.3\% to 45.3\%
for \texttt{qwen3.6} and from 1.4\% to 25.9\% for \texttt{mistral}
(Table~\ref{tab:exp2}).

Paired exact McNemar tests confirm these increases. From P1a to P2a,
\texttt{qwen3.6} has 25 new disclosures and none in the reverse direction
($p<0.001$); \texttt{mistral} has 16 and none
($p<0.001$). From P1b to P2b, the counts are 33 versus 1 for
\texttt{qwen3.6} ($p<0.001$) and 15 versus 1 for
\texttt{mistral} ($p<0.001$).

For example, Xenia asks about Astrid,
\emph{``What do they prefer when it comes to how they spend their time?''}
In both P2a and P2b, \texttt{qwen3.6} says that Astrid works long hours on a
farm and watches television shows. Both are latent Astrid attributes that
never appeared in the conversation.

\subsection{Effect of Profile Access on the Person Discussed}\label{sec:exp2-person}

With only the speaker's profile, wrong-person responses occur on 29.8\% of
turns for \texttt{qwen3.6} and 28.8\% for \texttt{mistral}, while right-person
responses occur on 14.5\% and 14.0\%. With both profiles, wrong-person
responses fall to 6.5\% and 8.6\%, while right-person responses rise to
56.2\% and 39.1\% (Table~\ref{tab:exp2}).

Paired exact McNemar tests confirm both changes. For right-person responses,
\texttt{qwen3.6} loses 7 and gains 92 responses ($p<0.001$), while
\texttt{mistral} loses 11 and gains 61 ($p<0.001$). For
wrong-person responses, \texttt{qwen3.6} stops 49 and starts 3
($p<0.001$), while \texttt{mistral} stops 45 and starts 4
($p<0.001$).

Our prompts never mention privacy, so these rates could mean either that the
agent ignores such a rule or that it was never given one. We re-ran the
cross-reference turns on \texttt{qwen3.6} with two added system-prompt lines: a
privacy line telling the agent to use only what a person has said aloud, and a
strict line stating that each profile is confidential to its owner and that the
agent should say it does not know when asked about something unsaid. On the 80
turns asking about a latent attribute, disclosure falls from 47.5\% to 24.4\%
(privacy; $n=78$, exact McNemar $p<0.001$) and to 8.8\% (strict;
$n=80$, $p<0.001$). The 40 turns asking about a revealed attribute
measure what these rules cost, because there the answer is already public. The
privacy line leaves those answers largely intact, 61.5\% against a baseline of
75.0\% ($n=39$, $p>0.05$), whereas the strict line answers only 20.5\%,
refusing 22 of 39 questions it had previously answered from public information
($p<0.001$). Over-disclosure is therefore a default rather than a refusal,
and the wording of the rule decides whether the agent withholds only what was
never said aloud or stops discussing the other person altogether.

\subsection{Response Quality}

Quality also increases when both profiles are available. For
\texttt{qwen3.6}, the mean rises from 2.26 (SD=1.04) with no profiles to
2.88 (SD=1.05) with both profiles ($p<0.001$,
rank-biserial $r=.28$). For \texttt{mistral}, it rises from 2.37
(SD=0.96) to 2.78 (SD=1.05) ($p<0.001$, $r=.18$).
Mann--Whitney tests are used because different replies were sampled across
conditions. Both effects are small.

\subsection{Effect of Spatial Information on Response Content}
\label{sec:exp2-gaze}

Finally, we compare each condition with and without the spatial line.
For right-person responses, P1a$\rightarrow$P1b changes are
13.9$\rightarrow$15.7\% for \texttt{qwen3.6} ($p>0.05$) and
17.3$\rightarrow$10.6\% for \texttt{mistral} ($p>0.05$).
For P2a$\rightarrow$P2b, the changes are 54.5$\rightarrow$57.6\%
($p>0.05$) and 36.0$\rightarrow$38.0\% ($p>0.05$).

No wrong-person or disclosure comparison is significant
(all $p>0.05$). Self-directed mix-ups
remain zero with and without spatial information.

Quality also shows no pattern that repeats across conditions and models.
The only significant comparison is Mistral P1
(2.58$\rightarrow$2.23, $p<0.01$); this change does not repeat in P2 or
for \texttt{qwen3.6}. We therefore find no consistent evidence that adding
spatial information changes response content.

\section{Human Validation}
\label{sec:validation}
 
Four annotators, blind to the answer key and condition, completed five tasks
(125 items); we report accuracy and, separately, agreement among annotators
(Fleiss' $\kappa$). On 35 text-only traps, accuracy was no better than chance
(54.5\% vs.\ 50\%, $p>0.05$; $\kappa=.134$), as intended: if the wording
revealed the addressee, annotators would be both correct and consistent. On
25 different, congruent traps with the spatial line, accuracy rose to 71.6\%
($p<0.001$) while agreement stayed low ($\kappa=.101$): the line moved
annotators toward the correct answer, not toward a shared habit. Annotators
confirmed 95.0\% of 10 authored contrasts as real conflicts ($\kappa=.865$)
and agreed with the attribute extractor, which underlies Experiment~2's
measures, on 91.0\% of 25 judgments ($\kappa=.799$). For the 26 of 30 replies
with a 3-of-4 majority on profile use, that judgment was associated with the
automatic quality score (Kruskal--Wallis $H=7.26$, $p<0.05$).
\section{Discussion}
\label{sec:discussion}

\subsection{RQ1: Spatial Cues Can Dominate Addressee Decisions}

RQ1 asked what happens when speaker orientation conflicts with the assigned
addressee. The models used orientation well when the two agreed, but followed
it on more than 85\% of incongruent turns. Warning \texttt{qwen3.6} that
orientation could be misleading reduced this behavior only modestly.

Prior work shows that gaze or head orientation can help identify an
addressee when combined with language and conversational context
\cite{jovanovic2006addressee,addressee_nonverbal2011}. Our results show why agreement
alone is not enough to test how these cues are used: when orientation and the
assigned addressee agree, simply following orientation gives the correct
answer. The incongruent cases in \textsc{LookAway} show that the tested models
often follow orientation when the two disagree. A likely reason is that trap wording is ambiguous by design, while the spatial line names a target outright (``looking at you''), so it is the clearest signal in the prompt; the warning results show the models keep following it even when told it may be wrong. Shared agents should therefore treat orientation as one cue rather than as the answer.

\subsection{RQ2: Profile Access Changes What the Agent Says}

RQ2 asked how access to multiple profiles affects responses about a
co-present person. Providing both profiles made the models more likely to
respond about the person being discussed and less likely to answer about the
questioner. However, it also increased the use of profile attributes that had
not appeared in the conversation. Identity mix-ups were rare, showing that
keeping users' identities separate does not by itself limit when their
profile information is used.

Shared agents may therefore need to distinguish information revealed in the
shared conversation from information available only in a user's profile.
This is consistent with work showing that appropriate information use depends
on the context in which information is shared
\cite{nissenbaum2004privacy,mireshghallah2024can}.

Our main prompts did not label latent attributes as private. The prompt
ablation in Section~\ref{sec:exp2-person} shows that an explicit rule lowers disclosure sharply
but does not remove it, so these results describe a default behaviour that
instruction suppresses rather than eliminates.

\subsection{Implications for Shared Multi-User Agents}

More context alone is not enough: orientation helps identify the addressee but
overrides it when the two conflict, and profiles help the agent answer about
the right person but bring in unrevealed information. Shared agents must
decide not only what information is available, but when and how strongly to
use it. These tests required the virtual setting, since conflicting
orientation is rare in natural recordings and hard to measure from video, but
can be set exactly in a 3D virtual environment, making it a practical testbed
for multi-user LLM agents.

\section{Limitations and Future Directions}
\label{sec:limitations}

\textsc{LookAway} uses synthetic, predefined speaker orientation rather than
measured eye gaze, model-generated conversations, and two-person interactions.
The main analyses use two open-weight models after the third model failed the
baseline addressee check. Future work should test other models, groups of more than two people, where the agent must choose among several possible addressees,
natural conversations, and noisy head- or eye-tracking signals.

Future studies should test the privacy rules on other models and conditions, including rules that give the agent an explicit record of what each person has said aloud, to see whether disclosure can be reduced without the over-refusal we observed under the strict rule. They should also test interactions in which addressee intent and speaker orientation come from real participants.

\section{Conclusion}
\label{sec:conclusion}

This work shows that shared LLM agents face two related challenges: deciding
when they are being addressed and deciding what information about co-present
users to use. In \textsc{LookAway}, the tested models relied strongly on
speaker orientation when it conflicted with the intended addressee, while
access to multiple user profiles improved person grounding but increased the use of information not revealed in the shared conversation, which could invade the privacy of the person being discussed. Together, these
results show that providing more context is not enough: shared agents must
also determine how strongly each cue should influence a decision and when
user-specific information is appropriate to use. 

\section*{Acknowledgment}

The authors used AI assistants to help write code, produce figures, and edit the text. The authors reviewed all AI-assisted output and take full responsibility for the content.

\bibliographystyle{IEEEtran}
\bibliography{references}

\end{document}